          [2001/04/25 1.1 (PWD)]
\documentclass[a4paper,twocolumn]{esapub} % European paper
\usepackage{graphicx}
\usepackage{natbib}

\def\lsim{\lower.5ex\hbox{$\; \buildrel < \over \sim \;$}}
\def\gsim{\lower.5ex\hbox{$\; \buildrel > \over \sim \;$}}
\title{High-energy emission from the stellar wind collisions in
$\gamma^2$ Velorum}
\author{Vincent Tatischeff}
\affil{Centre de Spectrom\'etrie Nucl\'eaire et de Spectrom\'etrie de Masse, 
IN2P3-CNRS and Universit\'e Paris-Sud, 91405 Orsay, France,
tatische@csnsm.in2p3.fr}
\author{Regis Terrier}
\affil{Astroparticule et Cosmologie, CNRS and Universit\'e Paris-VII, 
11 place M. Berthelot, 75005 Paris, France, rterrier@in2p3.fr}
\author{Fran\c{c}ois Lebrun}
\affil{DAPNIA, Service d'Astrophysique, CEA/Saclay, 91191
Gif-sur-Yvette, France, flebrun@cea.fr}

\begin{document}

%\keywords{}

\maketitle

\begin{abstract}
The binary system $\gamma^2$ Velorum (WC8+O7.5) contains the nearest known
Wolf-Rayet star to the Sun, at a distance of 258$_{-31}^{+41}$~pc. Its strong 
radio emission shows evidence for a partially absorbed nonthermal component, 
which has been interpreted as synchrotron emission from electrons accelerated 
in the colliding wind region. Inverse Compton cooling of these electrons in 
the intense UV radiation field from the O-type companion star could produce a 
significant hard X-ray and $\gamma$-ray emission, whose flux depends on the 
ratio of the energy densities of magnetic to seed photon fields. The Vela 
region was observed with the {\it INTEGRAL} satellite in 2003, as part of 
the Core Programme. No signals from $\gamma^2$ Vel are detected in the
images obtained with the IBIS/ISGRI coded aperture instrument in the
energy ranges 20--40 and 40--80 keV. From the derived 3$\sigma$ upper
limits, we show that the average magnetic field near the region of
stellar wind collision should be relatively high, $B$$\gsim$1~G. The
high-energy emission of $\gamma^2$ Vel might be detected with 
the forthcoming {\it GLAST} experiment.
\end{abstract}

\section{Introduction}

Radio continuum measurements of Wolf-Rayet (WR) stars have revealed at least 
ten sources with nonthermal (NT) emission, which presumably originates
from an interaction between the WR stellar wind and the
wind from a massive companion star (Chapman et al. 1999, and
references therein). The observed NT radio spectra and 
luminosities are well explained as synchrotron emission from relativistic 
electrons accelerated at strong shocks in the colliding wind region 
(Eichler and Usov 1993). 

Benaglia and Romero (2003) have recently studied the production of 
$\gamma$-ray emission in the binary systems WR~140, WR~146 and WR~147. They 
showed that inverse Compton (IC) scattering of the accelerated electrons in 
the strong photon fields of the early-type stars could produce $\gamma$-ray 
fluxes above the {\it INTEGRAL}/IBIS continuum sensitivity. In the case of 
WR~140, they also showed that the expected high-energy emission can account 
for the unidentified EGRET source 3EG J2022+4317. 

In the present paper, we consider the relatively close (i.e. with short 
orbital period $P_{orb}$=78.53~days) binary system $\gamma^2$ Velorum 
(=WR~11, van der Hucht 2001). With a distance of 258$_{-31}^{+41}$~pc, as 
determined from {\it Hipparcos} parallax measurements, this WC8+O7.5 binary 
contains the nearest WR star to the Sun. Radio and millimetre observations of
$\gamma^2$ Vel essentially revealed the strong thermal emission from the WR
ionized wind (e.g. Leitherer et al. 1997). However, Chapman et al.
(1999) found a significant steepening of the radio spectral index
between 3 and 20~cm, which they interpreted as evidence for a highly
attenuated NT component originating in the colliding
wind region, well within the radio photosphere of the WR star (Chapman et 
al. 1999). This contrasts with the wider binary systems studied by
Benaglia and Romero (2003), for which the free-free absorption of the
NT radio emission by the WR wind is less important. 

The existence of strong shocks in the colliding wind region of
$\gamma^2$ Vel is also supported by X-ray observations (Skinner et al.
2001, and references therein). In particular, {\it ROSAT} and {\it
ASCA} data have revealed a hot plasma emission ($kT$$\gsim$1~keV) with a
strong phase-locked variability, which was interpreted as a colliding wind 
shock emission originating deep within the dense and opaque WR wind, and 
showing significantly less photoelectric absorption at orbital phases when 
the cavity around the O-type companion star crosses the line of sight (Rauw 
et al. 2000, and references therein).        

The model for the high-energy emission from the stellar wind
collisions in $\gamma^2$ Vel is described in the next section. In Sect.~3,
we present the {\it INTEGRAL}/IBIS observations. The results are
discussed in Sect.~4. 

\section{High-Energy Emission Model}
\subsection{The Stellar Parameters and the Geometry of the System}

The geometry of the region of stellar wind collision in a close binary 
is described by Eichler and Usov (1993). Assuming the collision of two 
spherical winds, the distances $r_{WR}$ and $r_O$ from the WR and O-type 
stars, respectively, to the region where the winds interact is
\begin{equation}
r_{WR}={1 \over 1 + \eta^{1/2}}D_{bin}~~~{\rm and}~~~r_O={\eta^{1/2} 
\over 1 + \eta^{1/2}}D_{bin}.
\end{equation}
Here $D_{bin}$ is the distance between the two stars and $\eta$
denotes the ratio of the wind momentum fluxes of the O star to the WR
star:
\begin{equation}
\eta={\dot{M}_O V^\infty_O \over \dot{M}_{WR} V^\infty_{WR}},
\end{equation}
where $\dot{M}_i$ and $V^\infty_i$ are the mass-loss rate and the wind
terminal velocity of the star $i$, respectively.

\begin{table}[b!]
  \begin{center}
    \caption{Stellar and orbital parameters of $\gamma^2$ Vel.}
    \vspace{1em}
    \renewcommand{\arraystretch}{1.2}
    \begin{tabular}[h]{ccc}
      \hline
      \hline
      Parameters & WR star & O star \\
      \hline
      \hline
$\dot{M}$ (M$_{\odot}$/yr)  & 9.3$\times$10$^{-6}$$^{(a)}$ & 1.83$\times$10$^{-7}$$^{(a)}$  \\
$V^\infty$ (km/s)  & 1550$^{(a)}$ & 2500$^{(a)}$ \\
\hline
$R$ (R$_{\odot}$) & 3.2$^{(a)}$ & 12.4$^{(a)}$ \\
log($L$/L$_{\odot}$) & 5.00$^{(a)}$ & 5.32$^{(a)}$ \\
$T_{eff}$ (K) & 57100$^{(a)}$ & 35000$^{(a)}$ \\
      \hline \\
      \end{tabular}
\vspace{-0.25cm     
    \begin{tabular}[h]{cc}
      \hline 
      \hline 
      Distance & 258 pc$^{(b)}$ \\
      \hline 
      Orbital period & 78.53 days$^{(b)}$ \\
      \hline 
      Semi-major axis & 1.21 AU$^{(c)}$ \\
      Eccentricity & 0.326$^{(c)}$ \\
      $\Rightarrow$ Binary separation $D_{bin}$ & 0.82--1.60 AU \\
      \hline 
      NT radio spectral index $\alpha$ & -0.5$^{(d)}$ \\
      S$_{4.8~\rm{GHz}}$ & 13.5 mJy$^{(d)}$ \\
      \hline
    \end{tabular}}
    \label{tab:table}
  \end{center}
{\vspace{-0.3cm} \noindent $^{(a)}$ De Marco et al. (2000); $^{(b)}$ van 
der Hucht (2001); $^{(c)}$ Schmutz et al. (1997); $^{(d)}$ Chapman et al. 
(1999).}
\end{table}

Stellar and orbital elements of $\gamma^2$ Vel are given in
Table~1. These parameters have been obtained after the revised
distance determination with the {\it Hipparcos} satellite (Schaerer et
al. 1997). From the derived mass-loss rates and wind terminal
velocities, we get from eq.~(2) $\eta$=0.032, such that the colliding wind 
region is very close to the surface of the O star (eq.~1): as $D_{bin}$ 
varies from 0.82 to 1.60~AU in this eccentric binary, $r_O$ is contained 
between 2.1$R_0$ (at periastron) and 4.2$R_0$ (at apastron), where 
$R_0$=12.4~R$_{\odot}$ is the radius of the O-type star. In this 
environment, electrons accelerated at the strong shocks formed by the 
colliding stellar winds, should produce, together with the observed NT 
synchrotron radiation, a significant hard X-ray and $\gamma$-ray emission 
by IC interactions with UV photons emitted by the O star. The flux of 
high-energy emission relative to the NT radio emission depends essentially 
on the ratio of the energy densities of magnetic to seed photon fields.     

\subsection{The Magnetic Field in the Stellar Wind Collision Zone}

However, the strength of the magnetic field near the colliding wind region 
is not well-known. Following Eichler and Usov (1993), it may be estimated 
from the expected geometry of the magnetic field in an outflowing stellar 
wind as    
\begin{equation}
B(r)=B^s \times \left\{ \hspace {-0.2cm} \begin{array}{ll}
\vspace{0.2cm}
\big({R \over r}\big)^3 & {\rm for}~R \leq r < r_A~{\rm (dipole),} \\
\vspace{0.2cm}
{R^3 \over r_A r^2} & {\rm for}~r_A < r < R {V^\infty \over
V^{rot}}~{\rm (radial),} \\
{V^{rot} \over V^\infty}{R^2 \over r_A r} & {\rm for}~r > R {V^\infty
\over V^{rot}}~{\rm (toroidal),}
\end{array}
\right.
\end{equation}
where $R$ is the radius of the star, $B^s$ its surface magnetic field,
$V^{rot}$$\sim$(0.1--0.2)$V^\infty$ its surface rotation velocity and
$r_A$$\sim$(1--3)$R$ the Alfv\`en radius. For $B^S_{WR}$$\sim$0.1--10~kG 
(e.g. Maheswaran and Cassinelli 1994) and $B^S_O$$\lsim$$100$ G (e.g. 
Charbonneau and MacGregor 2001), we obtain $B$$\sim$0.1--10~G in the 
NT emission zone. It is possible that the average magnetic field is
enhanced by the compression produced by the strong shocks in the
region of the stellar wind collision. 

\subsection{The Nonthermal Emissions}

In the vicinity of the O star, energy losses of the accelerated
electrons are mainly due to IC scattering. Coulomb and
Bremsstrahlung losses can safely be neglected, as well as synchrotron
losses if the average magnetic field $B \lsim 50$~G. In the steady
state approximation, the equilibrium spectrum of the fast electrons is
then given by
\begin{eqnarray}
N(E_e) = {1 \over \dot{E}_{IC}(E_e)} \int_{E_e}^{\infty} 
\dot{Q}(E_e')~~~~~~~~~~~~~~~~~~~\nonumber \\
\times \exp \bigg[ - \int_{E_e}^{E_e'} {dE_e'' \over 
t_{esc}\dot{E}_{IC}(E_e'')}\bigg] dE_e'~,
\end{eqnarray}
where $\dot{Q}(E_e)$ is the differential injection rate of 
electrons accelerated at the shocks, $t_{esc}$$\sim$$\pi r_O / c$ is the
average time spent by the fast electrons in the NT emission
region ($c$ is the speed of light) and
\begin{equation}
\dot{E}_{IC}(E_e) \cong {4 \over 3} \sigma_T U_O 
\bigg({E_e \over m_e c^2}\bigg)^2
\end{equation}
is the IC energy loss rate, with $\sigma_T$ the Thomson
cross section, $m_e$ the electron mass and 
\begin{equation}
U_O = {L_O \over 2 \pi c R_O^2} \big[1-\cos \{ \arcsin(R_O/r_O) \} \big]
\end{equation}
the energy density of radiation from the O star 
in the NT emission region. 

We assumed the source spectrum of the accelerated electrons to be of the form 
\begin{equation}
\dot{Q}(E_e) \propto E_e^{-s}~~~~{\rm for}~E_e < E_{max},
\end{equation}
where the maximum energy $E_{max}$=$\gamma_{max} m_e c^2$ is obtained from 
the condition that the IC loss rate not exceed the acceleration rate 
(Eichler and Usov 1993):
\begin{eqnarray}
\gamma_{max}^2\simeq3\times10^8\eta\bigg({V_{WR}^\infty \over 2\times10^8~{\rm
cm~s}^{-1}}\bigg)^2 \bigg({B \over G}\bigg) \nonumber \\
\times \bigg({D_{bin} \over 10^{13}~{\rm cm}}\bigg)^2 
\bigg({L_O \over 10^{39}~{\rm erg~s}^{-1}}\bigg)^{-1}.
\end{eqnarray}
The power-law spectral index $s$ is related to the one of the radio
synchrotron emission $\alpha$=-0.5 (Table~1) by the classical
formula $\alpha$=$-(s-1)/2$. We thus have $s$=2, which is appropriate for 
strong shock acceleration.    

\begin{figure}[t!]
\centering
\includegraphics[width=0.9\linewidth]{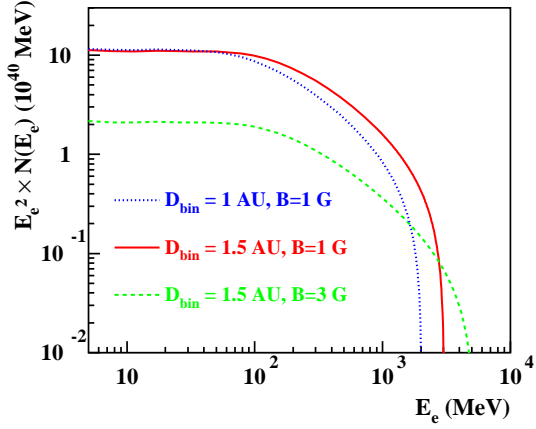}
\caption{Electron equilibrium spectra (times $E_e^2$) for 2 values of the 
binary separation $D_{bin}$ and average magnetic field $B$. All spectra are 
normalized to the production of the observed synchrotron flux density
S$_{4.8~\rm{GHz}}$=13.5 mJy (see Table~1).}
\end{figure}

Calculated electron equilibrium spectra are shown in Fig.~1. We see
that for reasonable values of the binary separation and average
magnetic field near the region of stellar wind collision, the electron 
energy distributions are cut off at a few GeV. In comparison, the
electrons accelerated in the wider binary systems WR~140, WR~146 and
WR~147 can reach energies of hundreds of GeV and thus produce IC
emission at higher energies than for $\gamma^2$ Vel (Benaglia and Romero 
2003).

\begin{figure}[t!]
\centering
\includegraphics[width=1.0\linewidth]{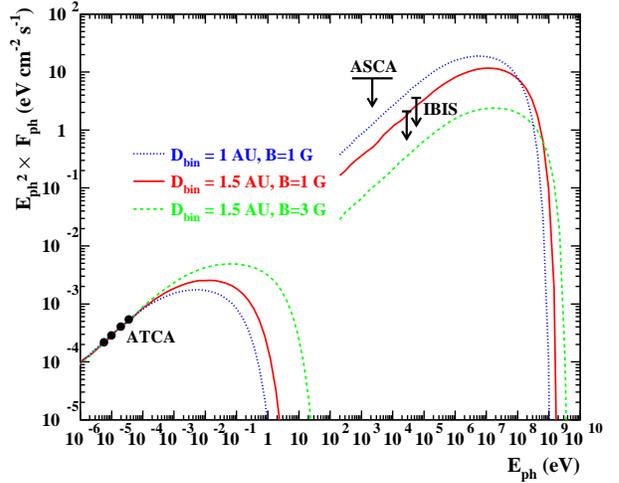}
\caption{Multi-band emission spectra (times $E_{ph}^2$) for
the same values of the binary separation $D_{bin}$ and average magnetic 
field $B$ as in Fig.~1. Also shown are the NT radio data at 3, 6, 13
and 20~cm obtained with ATCA (Chapman et al. 1999), the upper limit on
the 0.5--10 keV NT emission derived (see text) from ASCA observations 
(Rauw et al. 2000) and the {\it INTEGRAL}/IBIS upper limits (Sect.~3).}
\end{figure}

Multi-band emission spectra are shown in Fig.~2. Synchrotron and IC
radiations of fast electrons having the equilibrium spectra shown in
Fig.~1 were calculated from the detailed formalisms of Blumenthal 
and Gould (1970). The ASCA upper limit has been derived from the
observations of Rauw et al. (2000). These authors have detected a
relatively hard emission component with strong variability over the
binary orbital cycle, which they interpreted as the thermal emission 
from the shocked plasma in the colliding wind region. The intrinsic
(absorption-corrected) luminosity of this component in the 0.5--10 keV
energy range is $\gsim$10$^{32}$~erg~s$^{-1}$. Since the X-ray emission
does not show evidence for a NT component, we used this value to
derive the upper limit shown in Fig.~2. 

There is no point-like source in the third {\it CGRO}/EGRET catalog which is 
positionally coincident with $\gamma^2$ Vel (Hartman et al. 1999). But
because the energy of the accelerated electrons is limited to a few GeV, the 
predicted fluxes of the IC emission are exponentially cut off at 
$\sim$100~MeV, i.e. just above the low-energy threshold of the EGRET 
experiment. Furthermore, the detection of high-energy $\gamma$-ray emission 
from $\gamma^2$ Vel could be complicated by the proximity of the Vela 
pulsar, which is on time-average the brightest high-energy source in
the sky (the angular separation of $\gamma^2$ Vel and the Vela pulsar is 
4.9$^\circ$, whereas the FWHM of the EGRET point spread function at 100 MeV 
is $\sim$6$^\circ$). 
In fact, the third EGRET catalog contains 6 sources in the Vela region, 
which are almost certainly artefacts associated with the intense emission 
from the Vela pulsar (Hartman et al. 1999). This makes it very
difficult to obtain a statistically meaningful upper limit for the
high-energy IC emission of $\gamma^2$ Vel.

\section{INTEGRAL/IBIS observations}

INTEGRAL observations of the Vela region were performed in the
dithering mode of the satellite, on 2003 June 12--July 6 (fragmented
observations) and November 27--December 11. The total exposure time
amounts to $\sim$1.4$\times$10$^6$~s. The two observations 
were obtained when the binary was near apastron (from the ephemeris 
determined by Schmutz et al. 1997). During the two INTEGRAL observations, 
the average distance between the two stars was $D_{bin}$$\simeq$1.5~AU. In
comparison, the ATCA radio data were taken when the system was near
periastron, for an average binary separation $D_{bin}$$\simeq$1~AU. 

\begin{figure}[t!]
\centering
\includegraphics[width=1.0\linewidth]{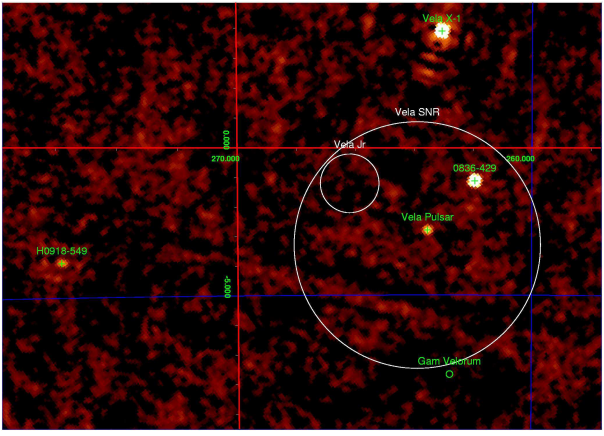}
\caption{IBIS/ISGRI significance map of the Vela region in the 20-40
keV energy band. The circles show the positions of the two supernova
remnants.}
\end{figure}

No signals from $\gamma^2$ Vel are detected in the images obtained with 
the IBIS/ISGRI coded aperture instrument in the energy ranges 20--40 keV
(Fig.~3) and 40--80 keV. The 3$\sigma$ upper limits obtained from the
instrument sensitivity are 7.4$\times$10$^{-5}$ and 
6.4$\times$10$^{-5}$~photons~cm$^{-2}$~s$^{-1}$ in the 20--40 and
40--80 keV energy ranges, respectively. 

\section{Results and Discussion}

The high-energy IC radiation of $\gamma^2$ Vel, which should accompany the
synchrotron radio emission revealed by the ATCA data, is not detected with
IBIS/ISGRI. A possible explanation is that the average magnetic field near 
the region of stellar wind collision is relatively high. Assuming that the 
intrinsic NT radio emission does not significantly vary over the orbital 
cycle, we obtain from the IBIS/ISGRI 3$\sigma$ upper limits $B$$\gsim$1~G 
(see Fig.~2). However, a significant variation of the NT radio emission is
observed in the well-studied binary WR~140. From 8 years of monitoring
the radio flux of WR~140 with the VLA, White and Becker (1995) have shown 
that the intrinsic NT component (before attenuation by free-free absorption)
seems to become weaker near periastron, which is not expected in the
model of spherically symmetric colliding winds. They proposed a new
model in which the WR wind is strongly enhanced in the star equatorial
disk. Such an effect could exist in the eccentric binary $\gamma^2$ Vel as 
well, and radio and $\gamma$-ray observations at the same orbital phases are 
required to specify the parameters of stellar wind collisions. 

The forthcoming $\gamma$-ray telescope {\it GLAST} might detect the
$\gsim$20~MeV IC emission of $\gamma^2$ Vel, provided that the proximity of 
the very bright Vela pulsar does not induce a contamination problem.
For an estimated 3$\sigma$ sensitivity
$S_{ph}$$\times$$E_{ph}^2$$\sim$0.4~eV~cm$^{-2}$~s$^{-1}$ at 100~MeV 
(for one year of observation and $\Delta E_{ph}$=$E_{ph}$), the IC
counterpart of the NT radio emission could be observed if the average 
magnetic field near the colliding wind zone is $B$$\lsim$10~G (see Fig.~2). 
In the case of a positive detection, it casts no doubt that $\gamma^2$ Vel 
would become a key object for understanding the processes of particle 
acceleration in binaries of early-type stars. 

\section*{Acknowledgments}

We acknowledge G. Bogaert and J. Kiener for useful discussions. 

% The following bibliography was produced with
%   \bibliographystyle{aa}
%   \bibliography{esapub}

\end{document}